\documentclass[preprint,aps,nofootinbib,tightenlines]{revtex4}

\usepackage{graphicx}

\usepackage{amsmath}
\usepackage{amsfonts}
\usepackage{amssymb}

\newcommand{\be}{\begin{equation}}
\newcommand{\ee}{\end{equation}}
\newcommand{\bear}{\begin{eqnarray}}
\newcommand{\eear}{\end{eqnarray}}
\newcommand{\beqstar}{\begin{eqnarray*}}
\newcommand{\eeqstar}{\end{eqnarray*}}

\newcommand{\tr}{{\rm Tr}}

\begin{document}


\vspace*{1.5cm} 
\title{$T$ Parity and the Littlest Higgs\vspace{0.5cm}}

\author{Ian Low}
\affiliation{Jefferson Physical Laboratory,
Harvard University, Cambridge, MA 02138
\vspace*{0.5cm}}

\begin{abstract}
\vspace*{0.5cm} 
We construct $T$-parity invariant extensions of the littlest
Higgs model, in which only linear representations of the full symmetry
group are employed, without recourse to the non-linear representations 
introduced by Coleman, Callan, 
Wess, and Zumino (CCWZ). These models are based on the symmetry
breaking pattern $SU(5)_l\times H_r / SO(5)$, where $H_r$ can be $SO(5)$ or
other larger symmetry groups. The structure of the models in the $SU(5)_l$
sector is identical to the littlest Higgs model based on 
$SU(5)/SO(5)$. Since the full symmetry group is realized linearly, these
models can be thought of as possible UV extensions of the $T$-invariant
model using non-linear representations via CCWZ, with whom they share
similar low energy phenomenology. We also comment on how to avoid
constraints from four-fermion operators on $T$-invariant models with or
without CCWZ construction. The electroweak 
data therefore place a very weak bound on the symmetry breaking scale,
$f \ge 450$ GeV.

\end{abstract}


\maketitle

\section{Introduction}
\label{sec:introduction}
With the Large Hadron Collider (LHC) at CERN setting to start running in 2007, we
are getting closer to unravel the mystery of the electroweak symmetry
breaking. In the standard model the electroweak symmetry is broken by the
vaccum expectation value of a scalar particle, the Higgs particle, whose
mass-squared receives quadratically divergent contribution from the UV
physics at the quantum level.
In order to stabilize the electroweak scale naturally, new physics is expected
at around 1 TeV, which is the energy scale that will be directly probed by
the LHC. Therefore much effort has been dispensed to construct models for new
physics at the TeV scale, which will be discernible by experiments in the
coming decade.

On the other hand, the standard model as we know today has been very
successful in confronting current experimental data. Precision measurements
in the last ten years reveal very little deviation from the prediction of
the standard model. In the low energies we can parametrize the effect of
new physics by a set of higher dimensional operators involving the standard
model fields only \cite{Weinberg:1978kz}, whose sizes are then constrained by
the precision electroweak measurements \cite{Buchmuller:1985jz}. The statement
that current data agree with the standard model predictions to a high
degree means the coefficients of these higher dimensional operators are
all very small. This in turn implies the scales of new physics, which
suppress the higher dimensional operators, are very large if one assumes
all the dimensionless numbers are order unity. The most loosely
constrained operators are those consistent with the (approximate)
symmetries of the standard model. Even for these operators current
experiments already suggest suppressions by energy scales as high as 5 - 10
TeV \cite{Barbieri:1999tm}. A naive 
interpretation would then be that new physics doesn't come in until 5 - 10
TeV, creating a tension with the naturalness principle which expects new
physics at 1 TeV.

There are two opposite attitudes we can take toward this ``little hierarchy
problem.'' One is to abandon the naturalness principle and try to make
peace with the fine-tuning. This approach has lead to the prediction of the
cosmological constant by the anthropic principle \cite{Weinberg:1987dv}, as
well as interesting models on physics beyond the standard model such as 
the split supersymmetry
\cite{Arkani-Hamed:2004fb}.  Nevertheless, naturalness consideration
as a principle has been successful in predicting the existence of new
physics in the past. One example is the self-energy of the electron (see,
for example, Refs.~\cite{Murayama:2000fm,nimatalk}), which classically is
linearly divergent due to the 
$1/r$ Coulomb potential. Naturalness principle then expects new physical
degree of freedom at the scale $m_e$, the electron mass. Indeed at the
energy scale $m_e$ we need to take into account the creation of
electron-positron pair out of the vaccum. After including the contribution
from the positron, the electron self-energy is only logarithmically
sensitive to the UV physics. Another example in which the naturalness
consideration works as expected is the low energy physics of QCD
\cite{nimatalk}. The charged $\pi$ mesons, being pseudo-Nambu-Goldstone bosons,
receive quadratically  divergent contribution to their mass from photons at
one loop. If the $\pi$ masses are to be in the 100 MeV range naturally, new
degrees of freedom should start showing up before 1 GeV. In this case, new
physics such as the $\rho$ and $a_1$  mesons do come in below the scale
suggested by naturalness. Therefore it is perhaps not surprising that the
second attitude, that is to take naturalness 
seriously, has been the driving force in theorizing physics
beyond standard model in the last couple decades.

Given that there are very few hints from experiments on the new physics at
the TeV scale, it is important to seek a resolution of the little
hierarchy problem without abandoning the naturalness principle. One simple
model-independent solution is the existence of a new $Z_2$
symmetry acting only on the new particles at the TeV scale
\cite{Wudka:2003se,Cheng:2003ju}. Examples of this $T$-parity
include the $R$-parity in the supersymmetric standard model and the
KK-parity in models with universal extra-dimensions. On the model building
side, a new class of theories dubbed the little Higgs theories was proposed
\cite{Arkani-Hamed:2001nc,Arkani-Hamed:2002qx,Arkani-Hamed:2002qy} with
solving the little hierarchy problem in mind, which stabilizes the 
electroweak scale naturally while raising the cutoff of the theory to 10 TeV, beyond
the probe of current electroweak data. 
There are many variants of the little Higgs models
\cite{Low:2002ws,Kaplan:2003uc,Chang:2003un,Chang:2003zn,Skiba:2003yf,Schmaltz:2004de},
as well as some examples of UV extensions above 10 TeV \cite{Katz:2003sn,Kaplan:2004cr}, 
and extensive phenomenological studies have been performed
\cite{Arkani-Hamed:2002pa,Hewett:2002px,Csaki:2002qg,Csaki:2003si,Gregoire:2003kr,Kilic:2003mq,Chen:2003fm,Kilian:2003xt,Yue:2004xt}.  
It turned out the impact of little Higgs models on the precision
observables is quite model dependent; some requires raising the cutoff
higher than 10 TeV, thus re-introducing the fine-tuning, and some do not. In
Refs.~\cite{Cheng:2003ju,Cheng:2004yc} it is shown that
it is possible to combine the idea of $T$-parity with the little Higgs
theories, and thus eliminate the precision electroweak constraints on a
large class of little Higgs models. Moreover, the lightest $T$-odd
particle, the LTP, is stable and massive in the desirable range to be a
good dark matter candidate.

There are in fact many ways to implement $T$-parity on little Higgs
models. In Ref.~\cite{Cheng:2003ju} a three-site moose model was
constructed in which the $T$-parity is a variant of the $Z_2$ reflection
symmetry between two sites. The third site then remains neutral under the
$T$-parity. Later it was realized that $T$-parity can be consistently
implemented on any non-linear sigma model based on a symmetric coset space
$G/H$. $T$-parity invariant models of the minimal moose type
\cite{Arkani-Hamed:2002qx} and the
littlest Higgs type \cite{Arkani-Hamed:2002qy} are
given  in which non-linear representations introduced by Callan,
Coleman, Wess, and Zumino (CCWZ) \cite{Coleman:sm,Callan:sn} are utilized
to assign the representations of the fermions \cite{Cheng:2004yc}. While from the
low energy perspective the formalism of CCWZ is the natural one to
consider, it makes no reference to what the possible UV extensions may
be. In this regard the three-site moose model, where all the matter is
assigned to linear representations of the full symmetry group $G$, is more
straightforward to UV-complete.\footnote{One trivial way, albeit not
favored by naturalness principle, is to simply complete to a linear sigma
model above 10 TeV.} Then the two-site minimal moose model can be considered
as a descendent of the three-site model by integrating out the neutral third site.

In this paper we consider extensions of the $T$-invariant models of the
littlest Higgs type, without recourse to the machinery of CCWZ, for which it
may be more straightforward to imagine possible UV completions \cite{jesse}. 
Similar to 
the minimal moose model, these extensions are achieved by extending the
global symmetry group, in this case to $SU(5)_l\times H_r / SO(5)_v$, where
$H_r$ is a group containing $SO(5)$ and the unbroken subgroup is the
vectorial $SO(5)$. In section II we consider the case with the minimal
group structure with $H_r=SO(5)$. Additional fermions need to be introduced
as well. A somewhat larger group structure with $H_r=SU(5)$ is considered
in section III, where less number of fermionic degree of freedom is
required. In both cases the $SU(5)_l$ sector is identical to the
$SU(5)/SO(5)$ littlest Higgs, thus distinguishing them from the minimal
moose type model where at least four link fields are necessary. Furthermore,
 the extra scalar and vector particles that come
with the extended global symmetry can all be
made heavy at around 10 TeV and integrated out of the spectrum. In the
last section we comment on the implications of our constructions on the
$T$-invariant models using non-linear representations of CCWZ and explain
how to avoid constraints from four-fermion operators on the model proposed 
in Ref.~\cite{Cheng:2004yc}.

\section{$SU(5)_l\times SO(5)_r / SO(5)_v$}


The non-linear sigma model here is based on the symmetry breaking pattern
$SU(5)_r \times SO(5)_r/SO(5)_v$, where $SO(5)_v = SO(5)_{l+r}$. Using the
same basis as in the littlest Higgs model in Ref.~\cite{Arkani-Hamed:2002qy}, we
write the generators in terms of $X_l^a$, $T_l^a$, and $T_r^a$, where
$X_l^a$ sit in the coset space $SU(5)_l/SO(5)_l$ and $T_i,\,
i=l, r$, sit in the $SO(5)_i,\, i=l, r$, respectively. The Lie algebra of
the full group looks schematically like
\begin{equation}
\label{lieso}
 [T_l^a, X_l^b]\sim X_l^c ,\quad  [X_l^a,X_l^b]\sim T_l^c , \quad
 [T_i^a, T_i^b] \sim T_i^c , \quad i = l, r.
\end{equation}
All other commutators are zero. The broken generators are $X_l^a$ and
$X_t \equiv T_l^a-T_r^a$ with the corresponding Goldstone bosons $\Pi_l = \pi_l^a X_l^a$
and $\Pi_t = \pi_t^a X_t^a$, which are parametrized as
\begin{equation}
\xi_o = e^{i\Pi_l/f} e^{i\Pi_t/f} \equiv \xi_l \xi_t \to g_l g_r\, \xi\, U ,
\end{equation}
where $g_l$, $g_r$, and $U$ belongs to $SU(5)_l$, $SO(5)_r$, and $SO(5)_v$,
respectively. Moreover, $U$ furnishes a non-linear representation of
$SU(5)_l\times SO(5)_r$, as was shown in CCWZ. Strictly speaking the coset
space $SU(5)_r \times SO(5)_r/SO(5)_v$ is not a symmetric space, however,
there is a $Z_2$ symmetry in Eq.~(\ref{lieso}) which can serve as the ascendent
of $T$-parity:
\begin{equation}
{\cal Z}_l : X_l \to - X_l, \quad T_i \to T_i , \quad i= l , r .
\end{equation}
Note that had we discarded the $SO(5)_r$ group, this would be the same $Z_2$
symmetry which makes it possible to 
impose $T$-parity on the littlest Higgs model in
Ref.~\cite{Cheng:2004yc}. The action of ${\cal Z}_l$ on $\xi_o$ gives
\begin{equation}
{\cal Z}_l (\xi_o) = \xi_l^\dagger \xi_t \to \tilde{g}_l\ g_r\ {\cal
  Z}_l(\xi_o) \ U,
\end{equation}
 where
${\cal Z}_l : g_l= e^{i(\epsilon_l \cdot T_l + \eta_l \cdot X_l)} \to
 \tilde{g}_l = e^{i(\epsilon_l \cdot T_l - \eta_l \cdot X_l)}.$
Taking the product of $\xi_o$ and ${\cal Z}_l(\xi_o)^\dagger$ we deduce that 
\begin{equation}
\xi_l^2 \to g_l\ \xi_l^2\ \tilde{g}_l^\dagger ,
\end{equation}
which then implies 
\begin{equation}
\label{xil}
\xi_l \to g_l\, \xi_l\, U_l^\dagger = U_l\, \xi_l \,\tilde{g}_l^\dagger .
\end{equation}
The element $U_l$ belongs to the $SO(5)_l$ group and furnishes a non-linear
representation of $SU(5)_l$ as in the $T$-invariant $SU(5)/SO(5)$ model.
Furthermore, given the vaccum expectation value in the littlest Higgs model
\begin{equation}
 \Sigma_0 = \left( \begin{array}{ccc}
                     &      & \openone \\
                     &   1  &          \\
           \openone  &      &     
                  \end{array}     \right),
\end{equation}
which satisfies $\Sigma_0 T_l \Sigma_0 = -T_l^T$ and $\Sigma_0 X_l \Sigma_0
= X^T_l$, $\tilde{g}_l$ has the property that $\Sigma_0\,
\tilde{g}_l^\dagger\, \Sigma_0 = g_l^T$. Therefore, if one defines $\Sigma_l= \xi_l^2
\Sigma_0$, it would transform as 
\begin{equation}
\label{sigmal}
\Sigma_l \to g_l\ \Sigma_l\ g_l^T .
\end{equation} 
That is, the $\Sigma_l$ transform 
just like the $\Sigma(x)$ in the littlest Higgs model and can be used to
write down the kinetic term for the Goldstone bosons in $\Pi_l$. On the
other hand, the kinetic term for $\Pi_t$ can be obtained using the 
object\footnote{Eq.~(\ref{xit2}) can be most easily deduced from Eq.~(\ref{xil}) 
and Eq.~(\ref{xm}), which is to be introduced later.}
\begin{equation}
\label{xit2}
\xi_t^2 \to U_l\, \xi_t^2 \, g_r^\dagger . 
\end{equation}
If we ignore the generators $X_l^a$, the coset space becomes $SO(5)_l\times
SO(5)_r / SO(5)$ 
and $\xi_t^2$ transforms in the familiar way $\xi_t^2 \to g_l\, \xi_t^2\,
g_r^\dagger$.

In the gauge sector, in the spirit of the three-site moose model in
Ref.~\cite{Cheng:2003ju}, three copies of $SU(2)\times U(1)$ are gauged:
\begin{eqnarray}
Q_{l1}^a &=& \left( \begin{array}{cc}
            \sigma^a/2 &     \\
                       & \phantom{\sigma^a/2}
               \end{array}            \right)_l,
 \quad \quad \ \ \phantom{aa} Y_{l1}= \frac1{10}\mbox{diag}(3,3,-2,-2,-2)_l , \\
Q_{l2}^a &=& \left( \begin{array}{cc}
            \phantom{\sigma^a/2} &     \\
                       & -\sigma^{a *}/2
               \end{array}            \right)_l,
 \quad \,  \,\, \phantom{aa} Y_{l2}= \frac1{10}\mbox{diag}(2,2,2,-3,-3)_l , \\
Q_r^a &=&  \left( \begin{array}{cc}
            \sigma^a/2 &     \\
                       & -\sigma^{a *}/2
               \end{array}            \right)_r,
 \quad \quad \ \ Y_{r}= \frac1{2}\mbox{diag}(1,1,0,-1,-1)_r . 
\end{eqnarray} 
Note that $Q_{l1}+Q_{l2}$ and $Q_r$ belong to $SO(5)_l$ and $SO(5)_r$,
respectively. The unbroken generators $Q^a$ and broken generators ${\cal
 Q}^a$ are thus
\begin{equation}
\label{gaugegenerator}
Q = Q_{l1} + Q_{l2} + Q_{r} ; \quad {\cal Q}_- = Q_{l1} - Q_{l2}; \quad
{\cal Q}_+ = Q_{l1} + Q_{l2} - 2 Q_{r}.
\end{equation}
The unbroken $SU(2)\times U(1)$ is naturally taken to be the electroweak
group $SU(2)_W\times U(1)_Y$.
The basis chosen in Eq.~(\ref{gaugegenerator}) is the eigenbasis of ${\cal Z}_l$ which interchanges
$Q_{l1}$ and $Q_{l2}$; $Q$ and ${\cal Q}_+$ are
even under ${\cal Z}_l$ whereas ${\cal Q}_-$ is odd. This
defines the $T$-parity for the gauge 
bosons, which forces the gauge couplings $g_{l1}=g_{l2}$. The heavy gauge
 bosons, ${\cal Q}_+$ and ${\cal Q}_-$, have masses 
of the order $gf$, which is typically at around 1 TeV. The $T$-even heavy
gauge bosons could in principle couple directly to the standard model matter,
which is also even under $T$-parity, and contribute dangerously to the
precision electroweak observables. This can be avoided by taking the gauge
coupling $g_r$ to be somewhat strong, right below $4\pi$, and
thus raising the mass of ${\cal Q}_+$ to be around 10 TeV. Then ${\cal
  Q}_+^a$ consist of mostly the gauge bosons in $Q_r^a$ and, if the
standard model matter is not charged under the strongly coupled gauge group
$Q_r^a$, their dangerous tree-level contributions to the precision
measurements are suppressed and hence safe. The gauge
sector here is completely similar to that of the three-site moose model with
$T$-parity in Ref.~\cite{Cheng:2003ju}.

With the gauge fields defined above and the transformation
properties given in Eqs.~(\ref{sigmal}) and (\ref{xit2}), we can write down the
scalar kinetic 
terms for the $\Pi_l$ and $\Pi_t$ 
\begin{equation}
\label{socovder}
{\cal L}_X  \supset \frac{f^2}8 \left( \tr|D_\mu \Sigma_l|^2 +  \tr|D_\mu
\xi_t^2|^2 \right),
\end{equation}
where
\begin{eqnarray}
\label{solder}
D \Sigma_l &=& \partial \Sigma_l- \sum_{j=1,2} \left\{ i g_{lj}
W_{lj}^a(Q_{lj}^a \Sigma_l + \Sigma_l Q_{lj}^{a T}) + i g_{lj}^\prime
B_{lj}(Y_{lj}\Sigma_l+\Sigma_l Y_{lj}^T) \right\}, \\
\label{sotder}
D \xi^2_t &=& \partial \xi^2_t  - v_{l}^a T^a_l\,  \xi^2_t + \xi^2_t \left(i
g_r W_r^a Q_r^a + i g_r^\prime B_r Y_r \right) .
\end{eqnarray} 
The object $v_{l \mu}^a T_l^a$ in $D_\mu \xi_t^2$ can be written compactly as
$v_{l \mu}^a T^a_l = \frac12(\xi_l^\dagger D_\mu \xi_l + \xi_l D_\mu
\xi_l^\dagger )$ which is defined in Ref.~\cite{Cheng:2004yc} and shown to
transform under the $U_l$ rotation like a gauge field:
$v_{l \mu}^a T_l^a \to U_l (v_{l \mu}^a T_l^a) U_l^\dagger + U_l
\partial_\mu U_l^\dagger$. In fact, the detailed form of $D_\mu \xi_t^2$
need not concern us here since all the uneaten Goldstone bosons residing
in $\xi_t^2$ will
be lifted to be massive at around 10 TeV and beyond the reach of precision
measurements. 

There are 14 Goldstone bosons in $\Pi_l$ and 10 Goldstone
bosons in $\Pi_t$. Their quantum numbers under the $SU(2)_W\times U(1)_Y$
are
\begin{eqnarray}
&&\Pi_l: \mathbf{\rm 1}_0 \oplus \mathbf{\rm 2}_{\pm 1/2} \oplus \mathbf{\rm 3}_0
\oplus \mathbf{\rm 3}_{\pm 1} , \nonumber \\
&& \Pi_t: \mathbf{\rm 1}_0 \oplus \mathbf{\rm 1}_{\pm 1} \oplus \mathbf{\rm
  2}_{\pm 1/2} \oplus \mathbf{\rm 3}_0 .
\label{soscalars}
\end{eqnarray}
The action of $T$-parity on the scalars is defined as
\begin{equation}
T: \Pi_l \to - \Omega \Pi_l \Omega, \quad \Pi_t \to \Omega \Pi_t \Omega ,
\end{equation}
where $\Omega = {\rm diag}(1,1,-1,1,1)$ ensures
that the Higgs doublet, which is taken to be the electroweak
doublet residing in $\Pi_l$, is even under $T$-parity.
Among the scalars in Eq.~(\ref{soscalars}), two copies of $\mathbf{\rm 1}_0\oplus
\mathbf{\rm 3}_0$, one from $\Pi_l$ and the other from $\Pi_t$, are eaten
by the heavy gauge bosons through Higgs mechanism. The complex triplet
in $\Pi_l$ become
massive through a potential radiatively generated at one loop due to the
gauge covariant derivative in Eq.~(\ref{solder}):
\begin{equation}
\label{sopotential}
 \sum_{j=1,2} \left\{ c_l g_{lj}^2 f^4 \sum_a \tr \left[(Q_{lj}^a
    \Sigma_l)(Q_{lj}^a \Sigma_l)^* \right]  + 
    c_l g_{lj}^{\prime 2} f^4\, \tr \left[(Y_{lj}\Sigma_l)(Y_{lj}\Sigma_l)^*
    \right] \right\},
\end{equation}
The remaining scalars are one complex
doublet and one complex singlet, both living in $\Pi_t$. 
They can be given a mass through the plaquette operator
\begin{equation}
\label{soheavymass}
 \epsilon_1 f^4\, \tr[\Omega^\prime \xi^2_t \Omega^\prime
(\xi_t^\dagger)^2] ,
\end{equation}
where $\Omega^\prime = {\rm diag}(1,1,-1,-1,-1)$. 
This plaquette operator is 
invariant under gauge transformations\footnote{For the purpose of verifying
gauge invariance, the symmetry breaking pattern can be thought of as
$SU(2)_{l1}\times SU(2)_{l2}\times SU(2)_r/SU(2)_v$. Then $\Omega^\prime$
commutes with $U_l$ when restricted to the gauge groups.} as well
as the $T$-parity. If $\epsilon_1$ is 
of order unity, the
scalars are massive at around 1 TeV. However, we can further raise the
scalar masses in $\xi_t$ to 10 TeV by making the plaquette strongly coupled,
$\epsilon_1 \sim 4\pi$. In so doing, the gauge and scalar sectors 
below 10 TeV is the same as in the littlest Higgs model, which has one set of
$SU(2)\times U(1)$ gauge bosons and one complex triplet scalar, all are
massive at
around 1 TeV, as well as one light Higgs doublet at the order 100 GeV.

The structure of this model in the $SU(5)_l$
sector is very similar to the $SU(5)/SO(5)$ littlest Higgs model. The Higgs
doublet and the triplet scalars are both from the $SU(5)_l$
sector
\begin{equation}
\Pi_l = \left( \begin{array}{ccc}
            \phantom{\sigma} & \frac{H}{\sqrt{2}} & \phi \\
             \frac{H^\dagger}{\sqrt{2}}  &   & \frac{H^T}{\sqrt{2}}  \\
            \phi^\dagger     &  \frac{H^*}{\sqrt{2}}  &   
             \end{array}  \right).
\label{tripletdef}
\end{equation}
So is the $T$-odd heavy gauge bosons, ${\cal Q}_- =
Q_{l1}-Q_{l2}$. All the scalars and gauge bosons associated with $SU(5)_r$
are made heavy at around 10 TeV. The lightness of the Higgs
doublet originates from the little Higgs mechanism working in the $SU(5)_l$
sector, in exactly the same way as the littlest Higgs model. The gauge
interaction in $\tr|D_\mu\Sigma_l|^2$ in Eq.~(\ref{socovder}) is identical
to that of the $\Sigma(x)$ in the littlest Higgs and
radiatively generates a potential, Eq.~(\ref{sopotential}),
giving the triplet mass, as well as the Higgs quartic coupling once the
triplet is integrated out \cite{Arkani-Hamed:2002qy}. The one loop
quadratic contribution to the Higgs mass-squared coming from the Higgs
quartic coupling is canceled by contributions from the triplet scalar
$\phi$ in Eq.~(\ref{tripletdef}); all
other scalars are not involved and can be as heavy as 10 TeV,
and hence decouple from the spectrum.

In the fermionic sector, the $T$-parity, which interchanges the two gauge groups
within $SU(5)_l$, forces identical fermion contents charged
under $SU(2)_{l1}$ and $SU(2)_{l2}$. A mirror fermion, charged under
$SU(2)_r$ which is neutral under $T$-parity, can be introduced to marry
one linear combination of the fermions charged under $SU(2)_{l1}$ and
$SU(2)_{l2}$. The remaining massless linear combination will be taken to be
the standard model doublet fermion, which does not carry any charged under the strongly
coupled gauge group $SU(2)_r$. Again this is reminiscent of the fermionic
sector in the three-site moose model in Ref.~\cite{Cheng:2003ju}. In order
to marry
fermions charged under $SU(5)_l$ and $SO(5)_r$, we need a function of
Goldstone bosons which
transforms as bi-fundamental under both groups and then introduce a Yukawa
type interaction. The function with the desired property is
\begin{equation}
\label{xm}
X_m \equiv \xi_l \xi_t^2 \to g_l X_m g_r^\dagger,
\end{equation}
which can be proven using CCWZ, who showed when one can and how
to construct a function of only Goldstone bosons with a desired transformation
property.
In the linear sigma model $X_m$ can be obtained from parametrizing
the Goldston bosons over the vaccum expectation value of a scalar
$\Phi \to g_l \Phi g_r^\dagger$.
A subtlety arises here because the physical Higgs doublet is contained in
$X_m$, which forces the mirror fermions to come in complete $SO(5)_r$
multiplet; otherwise a quadratic divergence to the Higgs mass would be induced through
the Yukawa-type interaction.
Therefore for each electroweak doublet in the standard
model, one introduces the following fermions:
\begin{equation}
\Psi_{l1}= \left( \begin{array}{c}
                     \psi_{l1} \\
                       0  \\
                       0  
             \end{array}    \right) ; \quad
\Psi_{l2}= \left( \begin{array}{c}
                       0 \\
                       0  \\
                      \psi_{l2}  
             \end{array}    \right) ; \quad
\Psi^c_r= \left( \begin{array}{c}
                     \psi^c_r \\
                       \chi^c_r  \\
                      \tilde{\psi}^c_r 
             \end{array}    \right) ; \quad
\Psi_r = \left( \begin{array}{c}
                     0  \\
                     \chi_r  \\
                     \tilde{\psi}_r
              \end{array}    \right) ,
\end{equation}
where $\Psi_{l1}$ and $\Psi_{l2}$ transform only under $SU(5)_l$, and
$\Psi^c_r$ and $\Psi_r$
transform only under $SO(5)_r$. Note that $\Psi_{l1}$ and $\Psi_{l2}$ have
the same charge under the diagonal gauge group $SU(2)_W\times
U(1)_Y$. Under $T$-parity, 
\begin{equation}
\label{sofermionT}
T: \Psi_{l1} \to - \Sigma_0 \Psi_{l2}, \quad \Psi^c_r \to -
\Psi^c_r , \quad \Psi_r \to - \Psi_r .
\end{equation}
The minus sign in the above serves to make the heavy fermions odd under
$T$-parity. Then the masses of the heavy, $T$-odd fermions could come from
the following interactions
\begin{equation}
\kappa_1 f \left( \bar{\Psi}_{l1} X_m \Psi^c_r + \bar{\Psi}_{l2} \Sigma_0
\tilde{X}_m \Psi^c_r \right) + \kappa_2 \Psi_r^T \Psi_r^c + h. c. ,
\label{sooddmass}
\end{equation}
where $\tilde{X}_m = \Omega \xi_l^\dagger \xi_t^2 \Omega $ is the image
of $X_m$ under $T$-parity. 
Eq.~(\ref{sooddmass}) gives a mass to the Dirac pair $(\psi_{l1}+\psi_{l2},
\psi_r^c)$, $(\chi_r, \chi_r^c)$, and $(\tilde{\psi}_r, \tilde{\psi}_r^c)$,
which are odd under the $T$-parity defined in 
Eq.~(\ref{sofermionT}). The standard model doublet is taken to be
the massless, $T$-even combination $\psi_{sm} \equiv (\psi_{l1} -
\psi_{l2})/\sqrt{2}$. We will see in the next section that for $H_r=SU(5)$ less
number of fermions 
is required and only $\psi_{l1}$, $\psi_{l2}$, and $\psi_r^c$ are necessary.


For Yukawa coupling, it suffices to focus on the top sector which generates
a large contribution to the Higgs mass. We write the third generation quark 
doublet as
$q^T = (b \ t) = (q_{l1}^T - q_{l2}^T)/\sqrt{2}$, where
$q_{lj}^T = ( b_{lj}\ t_{lj})$, for $j=1,2$. One also needs to introduce four
colored weak singlet fermions $t^\prime_{l1}$, $t^\prime_{l2}$, $t^{\prime
  c}_r$, and $t^{\prime c}$, and group them with the third generation quark in
the following fashion
\begin{equation}
Q_{l1} = \left( \begin{array}{c}
                      q_{l1}  \\
                      t^\prime_{l1} \\
                      0                      
                \end{array}  \right) , \quad
Q_{l2} = \left( \begin{array}{c}
                      0  \\
                      t^{\prime}_{l2} \\
                      q_{l2}
                \end{array}  \right) .
\end{equation}
The top Yukawa coupling is contained in the following interaction, summing
$i, j, k$ over 1, 2, 3 and $x, y$ over 4, 5, 
\begin{equation}
\label{yukawa}
\frac1{2\sqrt{2}} \lambda_1 f \epsilon_{ijk} \epsilon_{xy} \left[ (Q_{l1})_i (\Sigma_l)_{jx} (\Sigma_l)_{ky}
- (\Sigma_0\, Q_{l2})_i (\tilde{\Sigma}_l)_{jx} (\tilde{\Sigma}_l)_{ky}
\right ] u_3^{\prime c} 
 + \lambda_2 f t^\prime_+ t^{\prime c} + \lambda_3 f t^\prime_- t^c_r,
\end{equation}
where $t^\prime_{\pm} = (t^\prime_{l1} \mp
t^\prime_{l2})/\sqrt{2}$. The top Yukawa coupling here is the $T$-symmetried
version of the top Yukawa coupling in Ref.~\cite{Arkani-Hamed:2002qy}.
The image of $\Sigma_l$ under the $T$-parity, defined as $\tilde{\Sigma}_l$
in the above, is $\Omega (\xi_l^\dagger)^2 \Sigma_0 \Omega$. The
singlet fermions $t^\prime_-$ and $t^c_r$ are defined to be odd under
$T$-parity, whereas $t^\prime_+$, $t^{\prime c}$, and $u_3^{\prime c}$ are all $T$-even.
At leading order in the Higgs particle, Eq.~(\ref{yukawa}) gives
\begin{equation}
\lambda_1 (\sqrt{2} H q + f t^\prime_+) u_3^{\prime c} + \lambda_2 f
t^\prime_+ t^{\prime c} +
\lambda_3 f t^\prime_- t^c_r .
\end{equation} 
Therefore $t^\prime_-$ marries $t_r^c$ to become massive, and similarly for 
$t_+^\prime$ and  $\lambda_1 u_3^{\prime c} +  \lambda_2 t^{\prime c}$.
The remaining massless combination $u_3^c = (\lambda_2 u_3^{\prime c} -
\lambda_1 t^{\prime c})/\sqrt{\lambda_1^2+\lambda_2^2}$ has the desired
Yukawa coupling to $q$ with the strength $\lambda_t = \sqrt{2} \lambda_1
\lambda_2 /\sqrt{\lambda_1^2 + \lambda_2^2}$.

\section{$SU(5)_l\times SU(5)_r/SO(5)_v$}

In this section we enlarge $H_r$ to $SU(5)$ and show that less number of
fermions are needed for the construction. Similar to the case with $H_r=
SO(5)$, all the extra Goldstone bosons can be made heavy at 10 TeV and
integrated out of the spectrum. 
Written in terms of $T_i^a \in SO(5)_i$ and $X_i^a \in
SU(5)_i/SO(5)_i$, the Lie algebra 
of the full symmetry group is, again schematically,
\begin{equation}
\label{lie}
[T_i^a, T_i^b] \sim T_i^c , \quad [T_i^a, X_i^b]\sim X_i^c , \quad
[X_i^a,X_i^b]\sim T_i^c , \quad i=l,r.
\end{equation}
The broken generators in this case are $X_l^a$, $X_r^a$, and $X_t^a \equiv
T_l^a-T_r^a$. The Goldstone bosons $\Pi_l=\pi_l^a X_l^a$, $\Pi_r=\pi_r^a
X_r^a$, and $\Pi_t=\pi_t^a X_t^a$ can be parametrized as
\begin{equation}
\label{gaction}
\xi_u= e^{i\Pi_l/f}\ e^{i\Pi_r/f}\ e^{i\Pi_t/f} \equiv
\xi_l\ \xi_r\ \xi_t \to g_l\,g_r\, \xi\, U ,
\end{equation}
where $g_l$, $g_r$, and $U$ belongs to $SU(5)_l$, $SU(5)_r$, and $SO(5)_v$,
respectively.
Note that $[X_l^a, X_r^b]=0$ so $\xi_l \xi_r =\xi_r \xi_l$.
Eq.~(\ref{lie}) has several useful $Z_2$ symmetries:
\begin{equation}
{\cal Z}_l : X_l \to -X_l ;\quad {\cal Z}_r : X_r \to -X_r ; \quad 
{\cal Z}_t : X_l \to X_r ,\ T_l \to T_r.
\end{equation}
Taking the product of $\xi_u$ and ${\cal Z}_i(\xi_u)^\dagger$ and define
$\Sigma_i= \xi_i^2\Sigma_0, i=l, r,$ we deduce that 
\begin{equation}
\Sigma_i^2 \to g_i\ \Sigma_i^2\ g_i^T , \quad i=l,r ; \quad
\Sigma_t \equiv \xi_l \xi_r (\xi_t)^2 \xi_r^\dagger \xi_l^\dagger \to g_l\
\Sigma_t \ g_r^\dagger,
\end{equation}
where in obtaining the transformation property for $\Sigma_t$ we have focused
on the special case when the group action in Eq.~(\ref{gaction}) is restricted to
$g_l$. We can use the $\Sigma$s to write down the kinetic
terms for the Goldstone bosons: 
\begin{equation}
\label{covder}
{\cal L}_X  \supset \frac{f^2}8 \left( \tr|D_\mu \Sigma_l|^2 +  \tr|D_\mu
\Sigma_r|^2 + \tr|D_\mu \Sigma_t|^2 \right),
\end{equation}
where $D \Sigma_l$ is the same as in Eq.~(\ref{solder}) and
\begin{eqnarray}
\label{rder}
D \Sigma_r &=& \partial \Sigma_r- \left\{ ig_r W_r^a(Q_r^a \Sigma_r
+\Sigma_r Q_{r}^{a T}) + i g_r^\prime B_r (Y_r \Sigma_r + \Sigma_r
Y_r^T)\right\} ,    \phantom{\sum_{j}} \\
\label{tder}
D \Sigma_t &=& \partial \Sigma_t  + \Sigma_t \left(i
g_r W_r^a Q_r^a + i g_r^\prime B_r Y_r \right) - \sum_{j=1,2} \left( ig_{lj}W_{lj}^a
Q_{lj}^a + i g_{lj}^\prime B_{lj} Y_{lj} \right)\Sigma_t.
\end{eqnarray} 
In the above we have gauged the same generators as in the $H_r=SO(5)$ case.
Even though $\Pi_l$ and $\Pi_r$ are
contained in $\Sigma_t = \xi_l \xi_r (\xi_t)^2 \xi_r^\dagger
\xi_l^\dagger$, the interaction $\tr|D_\mu \Sigma_t|^2$ does not give rise
to extra kinetic terms for $\Pi_l$ and $\Pi_r$. 

There are 14+14+10 Goldstone bosons in total with the electroweak quantum
numbers 
\begin{eqnarray}
&&\Pi_l: \mathbf{\rm 1}_0 \oplus \mathbf{\rm 2}_{\pm 1/2} \oplus \mathbf{\rm 3}_0
\oplus \mathbf{\rm 3}_{\pm 1} , \nonumber \\
&&\Pi_r: \mathbf{\rm 1}_0 \oplus \mathbf{\rm 2}_{\pm 1/2} \oplus \mathbf{\rm 3}_0
\oplus \mathbf{\rm 3}_{\pm 1} ,  \\
&& \Pi_t\,: \mathbf{\rm 1}_0 \oplus \mathbf{\rm 1}_{\pm 1} \oplus \mathbf{\rm
  2}_{\pm 1/2} \oplus \mathbf{\rm 3}_0 . \nonumber
\label{allscalars}
\end{eqnarray}
That is, there are three real singlets, one complex singlet, three
complex doublet, three real 
triplets, and two complex triplets. The action of $T$-parity, which descends from
the $Z_2$ symmetry ${\cal Z}_l$, on the scalars is 
\begin{equation}
T: \Pi_l \to - \Omega_t \Pi_l \Omega_t, \quad \Pi_r \to \Omega_t \Pi_r \Omega_t,
\quad \Pi_t \to \Omega_t \Pi_t \Omega_t .
\end{equation}
Among the scalars in Eq.~(\ref{allscalars}), two copies of $\mathbf{\rm 1}_0\oplus
\mathbf{\rm 3}_0$ in $\Pi_l$ and $\Pi_t$ are eaten by the heavy gauge bosons as in the $H_r=SO(5)$
case. The two complex triplets 
in $\Pi_l$ and $\Pi_r$, as well as the complex doublet in $\Pi_r$, become
massive through a potential radiatively generated at one loop by the gauge interactions
in Eq.~(\ref{covder}):
\begin{eqnarray}
\label{potential}
&& \sum_{j=1,2} \left\{ c_l g_{lj}^2 f^4 \sum_a \tr \left[(Q_{lj}^a
    \Sigma_l)(Q_{lj}^a \Sigma_l)^* \right]  + 
    c_l g_{lj}^{\prime 2} f^4\, \tr \left[(Y_{lj}\Sigma_l)(Y_{lj}\Sigma_l)^*
    \right] \right\} \nonumber  \\
&& \quad \phantom{\sum_j } +  c_r g_{r}^2 f^4 \sum_b \tr \left[(Q_{r}^b
    \Sigma_r)(Q_{r}^b \Sigma_r)^* \right] + 
    c_r g_{r}^{\prime 2} f^4\, \tr \left[(Y_{r}\Sigma_r)(Y_{r}\Sigma_r)^*
    \right] ,
\end{eqnarray}
where the first line is generated by gauge groups sitting in $SU(5)_l$ and
gives a mass of order $g_{l}f \sim 1\ {\rm TeV}$ to the triplet in $\Pi_l$, while
the second line, generated by the gauge groups in $SU(5)_r$, gives masses
of order $g_r f \sim 10\ {\rm TeV}$ to every scalar in $\Pi_r$ except
the singlet.
The
scalars in $\Pi_r$ are at around 10 TeV because the 
gauge coupling is strong, $g_r\sim 4\pi$. The remaining scalars, which
do not obtain a mass through Eq.~(\ref{potential}), are two complex
doublets, residing in $\Pi_l$ and $\Pi_t$ respectively, one complex singlet
from $\Pi_t$, and one real singlet in $\Pi_r$. 
To ensure that there is
only one light electroweak doublet, which is the Higgs, we can put in the
following plaquette operators
\begin{equation}
\label{heavymass}
{\cal L}_X \supset 
 \epsilon_1 f^4\, \tr(\Omega^\prime \Sigma_t \Omega^\prime
\Sigma_t^\dagger) + \epsilon_2 f^4\, \tr (\xi_r^2 + \Omega \xi_r^2 \Omega) .
\end{equation}
If $\epsilon_i \sim 4\pi, i=1,2$, all the scalars, except for the doublet
and the triplet in $\Pi_l$, are massive at 10 TeV
and we have the same gauge and scalar sectors below 10 TeV as in the
littlest Higgs model.

In the fermionic sector, again we need to introduce identical fermion
contents charged under $SU(2)_{l1}$ and $SU(2)_{l2}$, and mirror
fermions charged under $SU(2)_r$. We will use $\Sigma_t$ to write down a 
Yukawa-type interaction to give the TeV fermion a mass. Here the mirror fermions
do not have to fill a complete multiplet of the $SU(5)_r$ because the induced 
quadratic divergence gives mass to only the doublet in $\Pi_t$,
but not the physical Higgs in $\Pi_l$:
\begin{equation}
\Psi_{l1}= \left( \begin{array}{c}
                     \psi_{l1} \\
                       0  \\
                       0  
             \end{array}    \right) ; \quad
\Psi_{l2}= \left( \begin{array}{c}
                       0 \\
                       0  \\
                      \psi_{l2}  
             \end{array}    \right) ; \quad
\Psi^c_r= \left( \begin{array}{c}
                     \psi^c_r \\
                       0  \\
                       0  
             \end{array}    \right),
\end{equation}
where $\Psi_{l1}$ and $\Psi_{l2}$ transform only under $SU(5)_l$ and $\Psi^c_r$
transforms only under $SU(5)_r$. Under $T$-parity,
\begin{equation}
\label{fermionT}
T: \Psi_{l1} \to - \Sigma_0 \Psi_{l2}, \quad \Psi^c_r \to -
\Psi^c_r .
\end{equation}
Then the masses of the heavy, $T$-odd fermions could come from
the following interactions
\begin{equation}
\kappa f \left( \bar{\Psi}_{l1} \Sigma_t \Psi^c_r + \bar{\Psi}_{l2} \Sigma_0
\tilde{\Sigma}_t \Psi^c_r \right),
\label{oddmass}
\end{equation}
where $\tilde{\Sigma}_t = \Omega (\xi_l^\dagger \xi_r \xi_t^2 \xi_r^\dagger \xi_l) \Omega$.
At one loop Eq.~(\ref{oddmass}) generates a
plaquette operators similar to the $\epsilon_1$ term in
Eq.~(\ref{heavymass}), which gives the doublet and singlet scalars in
$\Pi_t$ masses in the TeV range. The top Yukawa coupling can be written
down in a fashion identical to Eq.~(\ref{yukawa}) by introducing additional
singlet fermions. In the end there are fewer number of heavy fermions at 1 TeV
than in the $H_r=SO(5)$ case, while the numbers of gauge and scalar particles below
10 TeV in both cases are identical, which are the same as in the littlest Higgs model.

\section{Discussions}
\label{sec:con}

So far we have constructed extensions of the littlest Higgs model with
$T$-parity. In all cases encountered so far, whether we choose to utilize
CCWZ or not, the spectrum in the gauge and scalar sectors can be the same as in
the original littlest Higgs model; extra gauge bosons and scalar particles
can be lifted to be heavy at around the cutoff scale and integrated
out. One can ask 
the question whether it is possible to lift the mass of the mirror fermions
to 10 TeV as well and thus obtain a spectrum identical to
the littlest Higgs model. It turns out not possible to do so because the
Yukawa-type interactions giving masses to the mirror fermions,
Eqs.~(\ref{sooddmass}) and (\ref{oddmass}), contain vertices with one
standard model fermion, one mirror fermion, and a Goldstone boson which
look like $\kappa f \bar{\psi}_{sm} \pi^a \psi_r^c$. Such a vertex would, at one
loop through the box diagram, give a finite contribution to the
four-fermion operator $c_f (\bar{\psi}_{sm}\bar{\sigma}_\mu 
\psi_{sm})(\bar{\psi}_{sm} \bar{\sigma}^\mu \psi_{sm})$ with  
\begin{equation}
c_f \sim \frac{1}{16\pi^2}(\kappa f)^4 \left(\frac1{f^2}\right)^2 \frac1{(\kappa f)^2} =
\frac1{16\pi^2} \frac{\kappa^2}{f^2},
\end{equation}
where $(1/f^2)^2$ comes from the two Goldstone boson propagators and
$1/(\kappa f)^2$ from the mass term in the fermion propagators. The size of
such four-fermion operators is severely constrained by various experiments
to be $c_f \le 1/(5 - 15\ {\rm TeV})^2$ \cite{Giudice:2003tu}. If we take
$f$ to be its natural size $\sim 1$ TeV, the mass of the mirror fermion is
then constrained to be $\kappa f \le 0.8$ TeV. That is the heavy fermions
need to be slightly lighter than 1 TeV to cutoff the dangerous contribution to the
four-fermion operators.

Such constraints from the four-fermion operators are in fact not considered
in the original construction of $T$-invariant models using CCWZ in
Ref.~\cite{Cheng:2004yc}. It turns out that, because the standard model
fermions there have the kinetic term $\bar{\psi}(\xi^\dagger D_\mu \xi +\xi D_\mu
\xi^\dagger)\psi$, four-fermion operators are generated with unsuppressed
coefficients $1/f^2$, which would force 
$f$ to be in the order of 10 TeV and introduce fine-tuning to the Higgs
mass.\footnote{Correspondences with R. Rattazzi and R. Barbieri on this
  subject are gratefully acknowledged.} Nevertheless, the model with
$H_r=SO(5)$ in Section II inspires an alternative
construction using CCWZ, without extending the global/gauge symmetry group,
 which is safe from large four-fermion
operators. The construction proceeds as follows. In the $T$-invariant $SU(5)/SO(5)$ model in
Ref.~\cite{Cheng:2004yc} all the fermions are assigned to 
transform under the unbroken $SO(5)$ according to the prescription of CCWZ:
\begin{equation}
\psi= \left( \begin{array}{c}
                 \psi_1  \\
                 \chi    \\
                 \psi_2
             \end{array}   \right) \to U \psi,
\end{equation}
where $U$ belongs to the unbroken $SO(5)$ and non-linearly realizes the
full $SU(5)$ group. This assignment of fermion representation has the
advantage of getting rid of all the tree-level contributions to the
electroweak observables from new particles at 1 TeV. However, it also
results in the particular kinetic term giving rise to large
four-fermion operators. The remedy is to follow the idea of introducing
mirror fermions whose lightness serves to cutoff the size of the
four-fermion operators. That is, we will introduce two doublets
$\varphi_1$ and $\varphi_2$ transforming linearly under $SU(2)_1$ and
$SU(2)_2$ respectively. These two doublets are mapped into each other under
$T$-parity which interchanges $SU(2)_1$ and $SU(2)_2$. The mirror fermions,
which need to be neutral under $T$-parity, will be assigned to a complete
multiplet of the unbroken $SO(5)$ and transform non-linearly under $SU(5)$.
More specifically,
\begin{equation}
\Psi_1 = \left( \begin{array}{c}
                 \varphi_1  \\
                  0  \\
                  0
             \end{array}  \right) \to g_{u} \Psi_1 ; \quad
\Psi_2 = \left( \begin{array}{c}
                  0  \\
                  0  \\
                  \varphi_2
             \end{array}  \right) \to g_{u} \Psi_2 ; \quad
\Psi^c = \left( \begin{array}{c}
                 \varphi^c  \\
                  \chi^c  \\
                 \tilde{\varphi}^c
             \end{array} \right) \to U \Psi^c ,
\end{equation}
where $g_u$ is the $SU(5)$ rotation.
The mirror fermion $\Psi^c$
needs to be in complete $SO(5)$ multiplet; otherwise a two-loop quartic divergence 
would be generated \cite{Cheng:2004yc} and contribute dangerously to the Higgs mass.
 Again $T$-parity maps $\Psi_1\to
-\Sigma_0 \Psi_2$ and $\Psi^c \to - \Psi^c$. A Yukawa-type interaction
which gives the mirror fermion a mass is
\begin{equation}
\kappa f (\bar{\Psi}_1 \xi \Psi^c + \bar{\Psi}_2 \Sigma_0 \Omega
\xi^\dagger \Omega \Psi^c),
\end{equation}
where $\xi = e^{i\Pi/f} \to g_u\, \xi\, U^\dagger$. 
In this way the
standard model fermion $\varphi_{sm}=(\varphi_1-\varphi_2)/\sqrt{2}$ has the normal
kinetic term $\bar{\varphi}_{sm} D_\mu \varphi_{sm}$, whereas the mirror fermion
$\Psi^c$ has the CCWZ kinetic term $\bar{\Psi}_c(\xi^\dagger D_\mu \xi +\xi
D_\mu \xi^\dagger)\Psi_c$, which results in four-fermion operators consist of the
TeV fermions, but not the standard model fermions. In order to give masses
to $\chi^c$ and $\tilde{\varphi}^c$, we also need to introduce $\chi$ and
$\tilde{\varphi}$ which can sit in a complete spinor representation of
$SO(5)$ along with another singlet $\tilde{\chi}$, as discussed in
Ref.~\cite{Cheng:2004yc}. The Yukawa coupling for the top quark can be
written down in a way completely similar to Eq.~(\ref{yukawa}).

The extended models with $T$-parity discussed in Sections II and III, as
well as the modified CCWZ construction mentioned in the previous paragraph,
solve the little hierarchy problem naturally while at the same time are
consistent with data from precision measurements. 
After eliminating the constraint from the 
four-fermion operators, the strongest constraint on the $T$-invariant
models comes from the correction
to the $\rho$ parameter from the $T$-odd gauge bosons, which only restricts
the symmetry breaking scale $f$ to be larger than 450 GeV \cite{Cheng:2004yc}.
The introduction of
$T$-parity not only eliminates all the tree-level corrections to the
precision observables from new particles responsible for canceling the
quadratic divergence of the Higgs mass, but also adds a bonus of predicting
a weakly interacting massive particle, the lightest T-odd particle (LTP), which
is stable and serves as a dark matter candidate. The $T$-invariant
extensions of the littlest Higgs model, whether using CCWZ or not, all
share very similar phenomenology in that the $T$-odd particles need to be
pair-produced. Moreover, the LTP is most likely to be the $B^\prime$ gauge
boson, which is lighter than other $T$-odd particles because of the small
$U(1)$ gauge coupling and the large normalization factor in the hypercharge
assignment. Much of the discussion on the low energy phenomenology runs
parallel to those in Ref.~\cite{Cheng:2004yc}, since the spectra below 10
TeV are very similar, and will not be repeated.

\begin{acknowledgments}
Numerous discussions with Hsin-Chia Cheng are gratefully acknowledged, who
collaborated at various stages of this project and shared many important
insights. Helpful conversations with Nima Arkani-Hamed and Jesse Thaler are
also acknowledged. This work is supported in part by the National Science
Foundation under grant PHY-0244821. 
\end{acknowledgments}


\end{document}